\documentclass{aa}
\usepackage{graphicx}
\usepackage{latexsym,natbib}
\newcommand{\beas}{\begin{eqnarray*}}
\newcommand{\eeas}{\end{eqnarray*}}
\def\fig{Fig.\,}
\def\sec{Sect.\,}
\def\tab{Tab.\,}
\def\cf{cf.~}
\def\eg{e.g.~}
\def\eq{\!=\!}
\def\gtsim{~\rlap{\lower -0.5ex\hbox{$>$}}{\lower 0.5ex\hbox{$\sim\,$}}}

\def\magsqarcsec{mag$/$\raisebox{-0.4ex}{\hbox{$\Box^{\prime\prime}$}\,}}
\def\kms{km$\,$s$^{-1}$\,}
\def\halpha{$\mathrm{H}$\@{\sc$\alpha$}\,}
\def\hone{$\mathrm{H}$\,{\sc i}\,}
\def\htwo{$\mathrm{H}$\,{\sc ii}\,}
\def\s0{S$0$}
\def\rbr{\hbox{$R_{\rm br}$}\,}
\def\hin{\hbox{$h_{\rm in}$}\,}
\def\hout{\hbox{$h_{\rm out}$}\,}
\def\mubr{\hbox{$\mu_{\rm br}$}\,}

\def\rbrdhin{\hbox{\rbr$/$\hin}}

\begin{document}
\title{Outer edges of face-on spiral galaxies
\thanks{Based on observations obtained at the German-Spanish Astronomical 
Center (DSAZ), Calar Alto, jointly operated by the Max-Planck-Institut f\"ur 
Astronomie Heidelberg and the Spanish National Commission for
Astronomy.}
}

   \subtitle{Deep optical imaging of NGC\,5923, UGC\,9837, and NGC\,5434}

   \author{M.~Pohlen\inst{1,2} \and R.-J.~Dettmar\inst{1} \and R.~L\"utticke\inst{1} \and G.~Aronica\inst{1}}

   \offprints{M. Pohlen, \\\email{pohlen@ll.iac.es}}

   \institute{Astronomisches Institut der Ruhr-Universit\"at Bochum, 
D-44780 Bochum, Germany
    \and
Instituto de Astrof\'{\i}sica de Canarias, E-38200 La Laguna, Tenerife, Spain}

   \date{Received Month xx, xxxx; accepted Month xx, xxxx}
  \abstract{
We present deep optical imaging of three face-on disk galaxies together with a detailed description of the reduction and calibration methods used, in order to measure the intrinsic shape of their outer stellar edges.
Whereas it is now well accepted that disks of spiral galaxies are not infinite exponential beyond galactocentric distances of about 3-5 radial scalelengths, the genuine structure of the truncation region is not yet well known.
Our data quantitatively establish a smooth truncation behaviour of the radial surface brightness profiles and is best described by a two-slope model, characterised by an inner and outer exponential scalelength separated at a relatively well defined break radius. 
This result disagrees with the frequently assumed sharply truncated nature of the radial surface brightness profiles and implies the presence of stars and even star-formation beyond the break radius.
In addition, we do not find a strong influence of a nearby companion on the ratio of the break radius to the radial scalelength. Our results denote new observational constraints for the search of the physical explanation for these smooth disk truncations. 
   }
   \maketitle
%

\section{Introduction}
Since the pioneering work of \cite{devauc1,devauc2} and \cite{free} the 
radial surface brightness, tracing the stellar distribution, is empirically 
known to be well described by an exponential decline. The inner part 
can be dominated by a bulge, bar, or ring component, whereas the outer disk  
follows an exponential quite well with only minor deviations related 
to spiral arms.
However, \cite{vdk79} has already pointed out that the profiles appear to break off towards the outer parts.
This is later quantitatively confirmed using optical filters by \eg \cite{vdk82a}, \cite{bd}, \cite{pohlen1,pohlen2}, and \cite{dgkrwe}. 
Just recently, data by \cite{floridoNIR} for the first time suggest these truncations to be also present in NIR images.
These so called cut-off radii are practically only observed in edge-on galaxies due to the preferred line-of-sight integration.
While the existence of the truncations is now well accepted there are still two major unsolved problems.  \newline 
On the one hand we are still missing a unique physical interpretation to describe this observational phenomenon.
The proposed hypotheses span a rather wide range of possibilities. \Citet{vdk87} has deduced a connection to the galaxy formation process describing the truncations as remnants from the early collapse. \cite{ferg3}, for example, have proposed an evolutionary scenario represented by the viscous disk evolution models. \cite{kenni} has suggested a dynamical critical star-formation threshold, and \cite{batta} have proposed a connection with large scale galactic magnetic fields. \newline 
On the other hand the detailed structure of the radial surface brightness distribution of disk galaxies is still not well constrained, although already  \cite{naeslund}, \cite{byun}, \cite{dgkrwe}, and \cite{pohlenphd} have found that the truncations are not that sharp as thought before. 
While the former studies assume the truncation region to be exponential, \cite{floridoNIR} used a model independent deprojection method to measure and analyse the truncation curve (difference between measured profile and an infinitely exponential decline).
However, nearly all studies used edge-on galaxies to search for truncations, whereas only the profiles of face-on galaxies are practically independent of possible line-of-sight effects caused by the integration across a multicomponent disk and less affected by the disturbing dust influence. 
Observing face-on galaxies is therefore mandatory to study the detailed quantitative behaviour at the truncation region and to determine the scalelengths in an unbiased way. 
Up to now \cite{vdkshos82} and \cite{shos} have reported the only known quantitative values of cut-offs in the literature for face-on galaxies (NGC\,3938 and NGC\,628).
\Citet{vdk82a} have just found indications of a radial truncation for NGC\,5033 and NGC\,3198 and no evidence in NGC\,4258 and NGC\,5055, all being late type galaxies.
Although calling the observed feature near the edge of the disk a 'truncation' \cite{vdk82a} have already stated that ``the edges are not infinitely sharp (we usually derive an upper limit to the scalelength beyond $R_{\rm max}$ of about 1\,kpc)''.
In a subsequent paper \cite{vdk88} has stated that out of the 20 face-on galaxies observed by \cite{wev} only four (NGC\,2681, NGC\,4203, NGC\,5005, NGC\,6340) ---all being early type galaxies and therefore probably possessing a different disk formation history--- do not show any sign for a drop off as judged from the last three contours. He has also recognized that only NGC\,5371 shows a relatively sharp edge but without quoting detailed numbers.  
Explaining this in addition to the much lower surface brightness for face-on systems, with the problem of dealing with intrinsic deviations from the circular symmetry of disks which could be hidden by an azimuthally averaged profile. 
These asymmetries, such as spiral arms, especially from the young population should unavoidably contribute in less inclined systems.
All these results are based on photographic material and up to now there is no systematic follow up of truncations for face-on galaxies. Especially a detailed quantitative modelling of radial surface brightness profiles based on higher quality CCD imaging data is missing.
However, there are at least two recent papers in the literature which both have not found any sign for a truncation investigating two nearly face-on galaxies.
\cite{barton} ---with a sensitivity limit of $\mu^{\rm V}_{\rm lim}\eq29.9$\,\magsqarcsec--- have not found any evidence for a cut-off for the galaxy NGC\,5383 out to approximately three Holmberg radii. 
They have measured an exponentially falling disk with several distinct departures from the smooth decline (two spiral arms with one outside the Holmberg radius), but their fitting yields unsatisfactory results.
Nevertheless, in addition to a nearby companion (UGC\,8877) this galaxy unfortunately shows a rather complex morphological structure (type: PSBT3*P according to the RC3\footnote{Third Reference Catalogue of bright galaxies \citep[RC3:][]{rc3}.}) and has also been rejected by \cite{vdk88} in his study.
\cite{weiner} also have found no sign of a truncation out to ten scalelength for the NGC\,4123 (type: SBR5 + starburst according to NED\footnote{NASA/IPAC Extragalactic Database (NED).}) in their azimuthally averaged light profile.
These studies have in common that both target galaxies are strongly barred and show a rather peculiar overall morphology. 
The detailed structure of the truncation region can be expected to give important constraints for the physical nature of these outer stellar edges. In addition our analysis has consequential effects on the determination of the truncation as a structural parameter which is important to compare observational results with model predictions. Depending on the true physical nature these outer edges may contain fossil evidences imprinted by the detailed galaxy formation and evolutionary history. 
\section{Sample Selection \& Observations}
\begin{figure}
\includegraphics[width=5.6cm,angle=270]{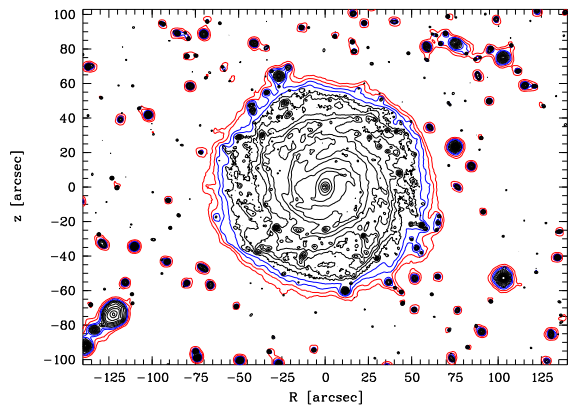}
\caption[]{Contour map ($\mu_{\rm JR}$) of NGC\,5923 from 26.9 to 18.0 equally spaced by 0.5 mag. \label{all_cont1}}
\end{figure}
\begin{figure}
\includegraphics[width=5.6cm,angle=270]{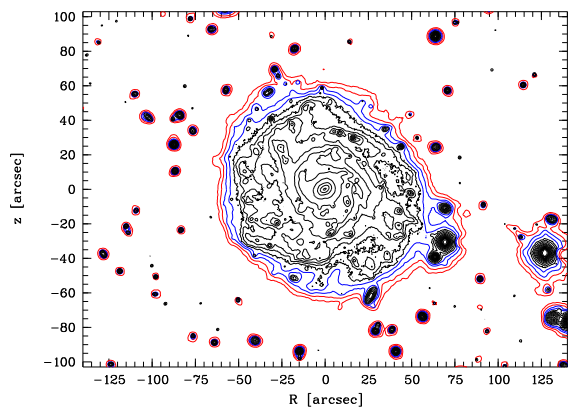}
\caption[]{Contour map ($\mu_{\rm JR}$) of UGC\,9837  from 26.8 to 18.0 equally spaced by 0.5 mag. \label{all_cont2}}
\end{figure}
\begin{figure}
\includegraphics[width=5.6cm,angle=270]{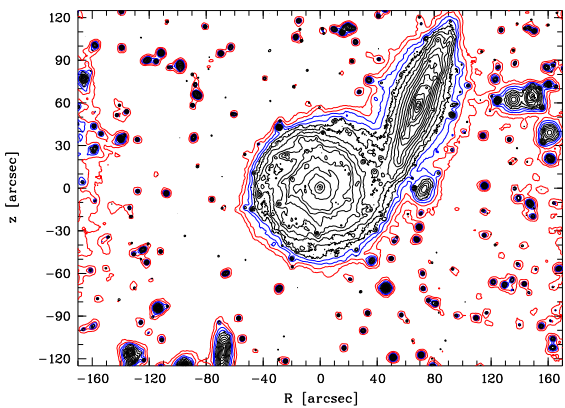}
\caption[]{Contour map ($\mu_{\rm JR}$) of NGC\,5434 from 27.0 to 18.0 equally spaced by 0.5 mag. \label{all_cont3}}
\end{figure}
\begin{figure}[tb]
\includegraphics[width=5.6cm,angle=270]{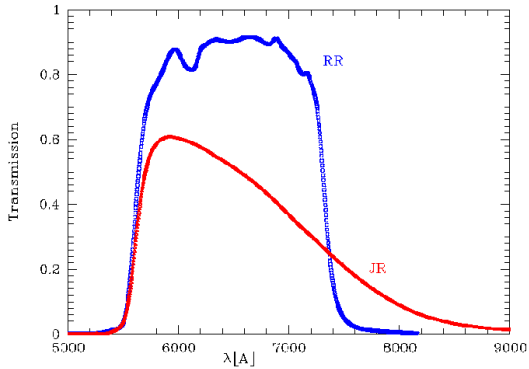}
\caption[]{R\"oser R (RR, \#648/168) and Johnson R (JR, \#641/158) filter curves. \label{CAHA_filt}}
\end{figure}

The main selection criteria for the observed objects are that they 
should be as close as possible to face-on disks. 
In a first step the galaxies are selected from the 
RC3 to be at least inclined less than 30\degr (deduced by $\cos(i)=a/b$).
According to the lack of a truncation in the profile of \s0-galaxies \citep[cf.][]{pohlenphd} early type galaxies $T\!<\!2$ are excluded as well as galaxies later than $T\!>\!6$ to have a well defined and symmetric stellar disk.
Additionally, we exclude galaxies classified as SB (including SABs) since a prominent bar may significantly disturb the assumed azimuthal symmetry while smoothing the profiles.
In order to obtain the selected galaxies with a decent resolution we first restrict the size to $D_{25}\gtsim2.0$\arcmin.
After visually inspecting all galaxies from the RC3 matching these criteria on DSS\footnote{Digitized Sky Survey (DSS)} images, our target galaxies should possess as few as possible foreground stars within the FOV (especially no bright ones) and should not show significant irregularities pointing to present or recent interactions.
It turns out that for a given observing period galaxies corresponding
to all the above mentioned selection criteria and good 
visibility for approximately $4$\,h during the night are rare.
Whereas from the visual inspection NGC\,4136 seems to be the 
ideal galaxy, technical flatfielding problems recognised at 
the telescope (described in \sec\ref{flatfielding})
forced us to choose galaxies with smaller diameters and we 
relaxed the size limitation to $D_{25}\gtsim1.8$\arcmin. 
Therefore we finally choose NGC\,5434, NGC\,5923, and UGC\,9837 
as good candidates (\cf\fig\ref{all_cont1}, \fig\ref{all_cont2}, and 
\fig\ref{all_cont3}).
NGC\,5923 and UGC\,9837 match all our criteria, despite some
intermediate bright stars in the FOV, whereas NGC\,5434 possesses 
two rather bright stars within the FOV.
Additionally, for NGC\,5434 there is an almost equal mass companion (NGC\,5434B; KPG 410) with similar velocity nearby.
Although NGC\,5434 is therefore probably interacting and even slightly overlapping with the companion (in projection), its shape is extraordinary regular and free of large scale asymmetries.
Owing to the lack of better suited galaxies, NGC\,5434 was observed to investigate effects of a possible ongoing interaction on the radial surface brightness of both galaxies.  
The global parameters of the observed face-on galaxies are shown in 
\tab\ref{F2sample}.
\begin{table*}
\begin{center}
{\normalsize
\begin{tabular}{ l  c c l  r@{.}l  c  c  c r@{.}l }
\hline
\rule[+0.4cm]{0mm}{0.0cm}
Galaxy
&RA 
&DEC
&RC3
&\multicolumn{2}{c}{T}
&Diam.
&$v_{\sun}$
&$v_{\rm vir}$
&\multicolumn{2}{c}{D} \\[+0.1cm]
&\multicolumn{2}{c}{(2000.0)}
&type 
&\multicolumn{2}{c}{}
&[\ \arcmin\ ]
&[\ \kms]
&[\ \kms]
&\multicolumn{2}{c}{[Mpc]} \\
\rule[-3mm]{0mm}{5mm}{\scriptsize{\raisebox{-0.7ex}{\it (1)}}}
&{\scriptsize{\raisebox{-0.7ex}{\it (2)}}}
&{\scriptsize{\raisebox{-0.7ex}{\it (3)}}}
&\hspace*{0.1cm}{\scriptsize{\raisebox{-0.7ex}{\it (4)}}}
&\multicolumn{2}{c}{{\scriptsize{\raisebox{-0.7ex}{\it (5)}}}}
&{\scriptsize{\raisebox{-0.7ex}{\it (6)}}}
&{\scriptsize{\raisebox{-0.7ex}{\it (7)}}}
&{\scriptsize{\raisebox{-0.7ex}{\it (8)}}}
&\multicolumn{2}{c}{{\scriptsize{\raisebox{-0.7ex}{\it (9)}}}} \\
\hline\hline \\[-0.2cm]
NGC\,5923 &152114.2&$+$414333 &.SXS4.. &4&0  &1.8 &5571 &5811.0& 80&7 \\
UGC\,9837 &152351.7&$+$580312 &.SXS5.. &5&0  &1.8 &2657 &2934.0& 40&8 \\
NGC\,5434 &140323.1&$+$092653 &.SA.5.. &5&0  &1.8 &4637 &4735.0& 65&8 \\
\hline
\end{tabular}
}
\caption[]{Global parameters of the observed face-on galaxies: 
{\scriptsize{\it (1)}} Principal name, {\scriptsize{\it (2)}} right ascension, and {\scriptsize{\it (3)}} declination,  {\scriptsize{\it(4)}} RC3 coded Hubble-type, {\scriptsize{\it(5)}} the Hubble parameter T, and {\scriptsize{\it(6)}}  the diameter in arcminutes \citep{rc3}. The {\scriptsize{\it(7)}} heliocentric radial velocities are taken from NED. According to the heliocentric radial velocities corrected for the Local Group infall onto Virgo {\scriptsize{\it(8)}} from LEDA, we estimated the {\scriptsize{\it (9)}} distances following the Hubble relation with the Hubble constant from the HST key project of $H_{0}\!=\!72$ km s$^{-1}$Mpc$^{-1}$ \cite[]{hst_h0}.
\label{F2sample} }
\end{center}
\end{table*}
The images were obtained during an observing run in May 2001 at the 
2.2\,m telescope of the Calar Alto observatory (Spain) equipped with CAFOS, providing a field-of-view of 16\arcmin\,using a SITe CCD with 15\,$\mu$m pixel size ($0.53$\,\arcsec pix$^{-1}$). 
Showing a nearly rectangular filter characteristic combined with a 
significantly higher peak efficiency, the so called R\"oser R 
(RR, \#648/168) filter is favoured to increase the signal-to-noise using  
the same integration time and was used instead of the standard 
Johnson R (JR, \#641/158) filter.
The different filter characteristics are shown in \fig{\ref{CAHA_filt}}.
In order to account for the crucial flatfielding problem of deep surface photometry, images for all possible flatfield strategies were taken: \mbox{dome-,}
\mbox{twilight-,} and nightsky-flatfields.
The nightsky-flatfields were acquired without loosing dark time 
by observing additional edge-on galaxies (covering less area
of the chip) and significant dithering of the rather small objects 
($D_{25}\!\approx\!2$\arcmin) across the large FOV ($\oslash$16\arcmin).
During each night several Landolt standard fields \cite[]{landolt} were 
observed to accomplish the photometric calibration.
\section{Image Reduction \& Photometric Calibration}
The 2048x2048 pixel SITe CCD chip (RON: 5.3\,e$^-$, GAIN: 2.7\,e$^{-}$/ADU) was windowed to improve the quite large read-out time (3min) of CAFOS. 
Approximately 400 pixels 
were excluded on each side vertical to the readout axis to 
preserve the overscan region. 
Additional 10 bias frames were taken at the beginning and end of 
each night. It turned out that there is a minor light leak within the 
CAFOS camera and bias frames taken with the dome
slit open are on average $\approx 0.2 - 2.9 $ADUs brighter, depending
on the actual brightness within the dome. 
Therefore all bias frames are inspected and according to our records rejected.
The following reduction procedures are performed with standard 
IRAF\footnote{Image Reduction and Analysis Facility: \newline 
\hspace*{0.5cm} http://iraf.noao.edu/} tasks.
The approximately 80 columns of the overscan region are averaged and 
an overscan vector is fitted with an 11th order polynomial,
using a $\pm 3\sigma$ rejection. This is removed row-by-row from 
each image by {\sl ccdproc}. 
A bias image is made for each night out of the remaining useful
10-15 overscan-subtracted bias frames with averaging and 
$\pm\!3\sigma$ rejection to remove cosmic rays by {\sl zerocombine}. 
The resulting bias images still show offsets of 
$\approx 4.7$ ADUs together with a large scale gradient and a large scale column and row pattern structure.
This master bias is than subtracted from all flat and object frames.
Dark frames taken during the day show that the dark current is 
equal or less than approximately 0.3 ADUs per 600 s exposure with no 
evident additional two-dimensional structure. 
It was not possible to reduce the light passing through the closed 
dome-slit to zero during the day. It is therefore not possible to 
decide if the remaining counts are either from real dark-current 
or from stray light passing into the camera, consequently no dark 
correction is applied. 
\begin{figure*}
\includegraphics[width=5.8cm,angle=270]{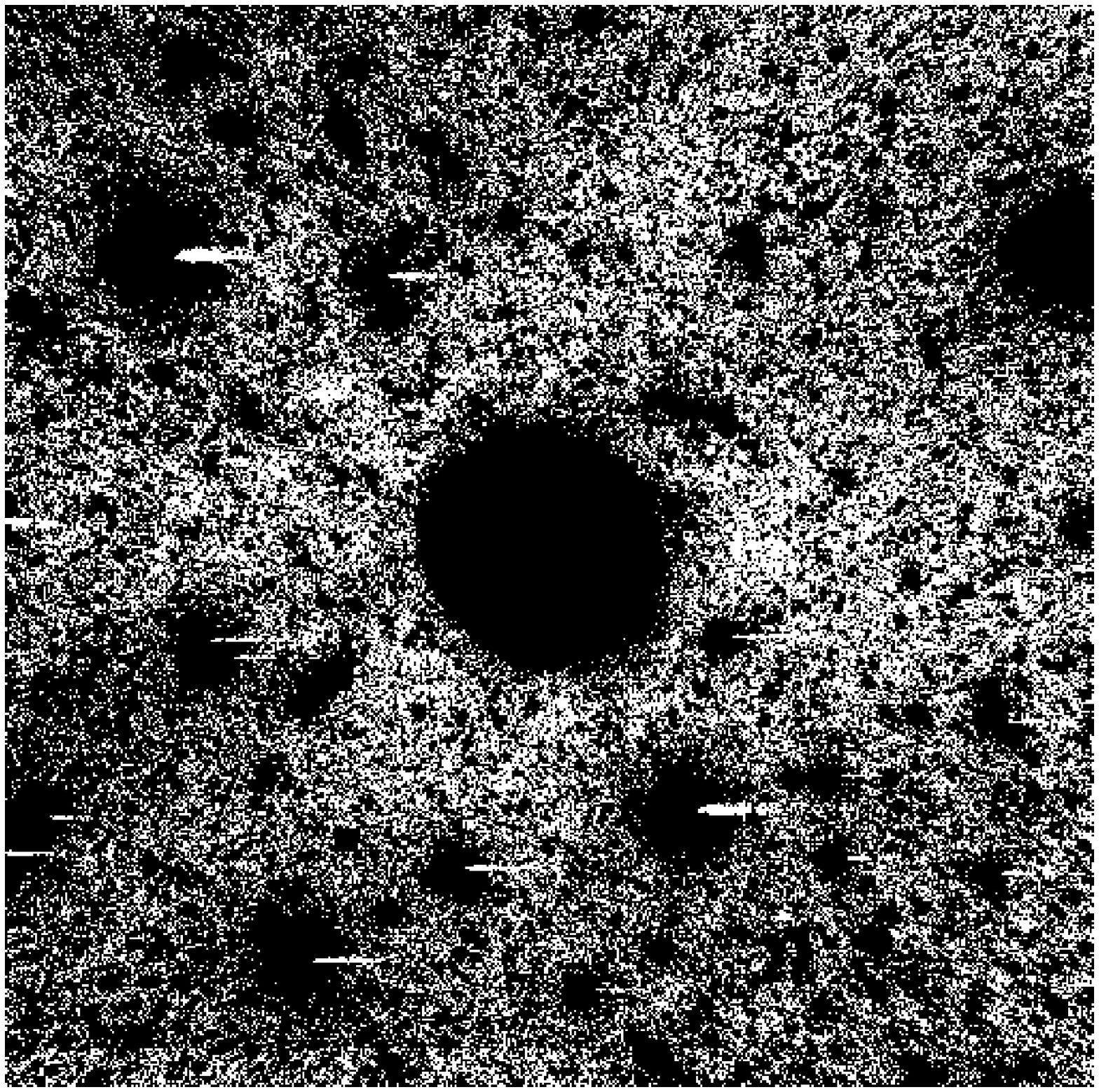}
\includegraphics[width=5.8cm,angle=270]{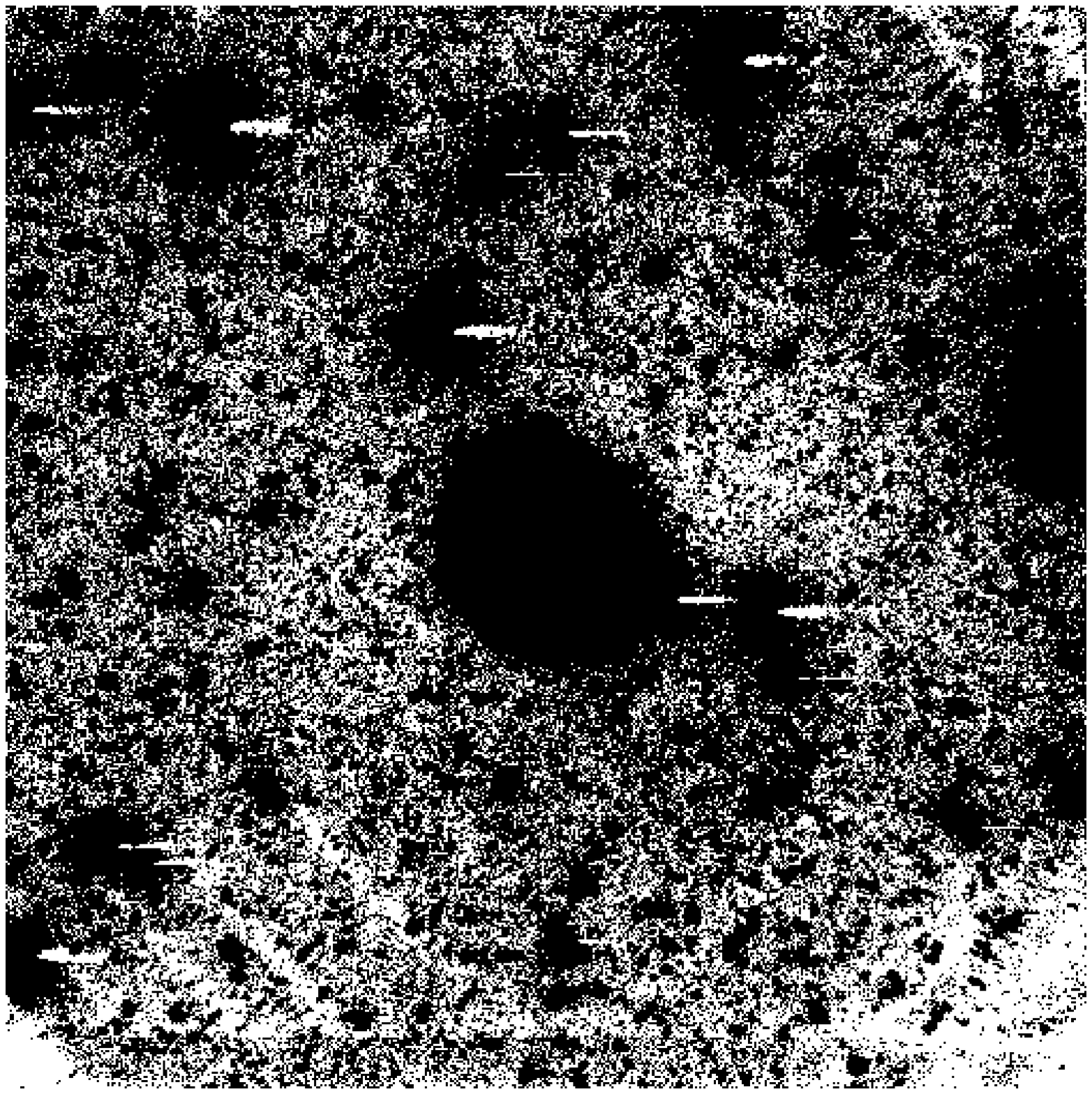}
\includegraphics[width=5.8cm,angle=270]{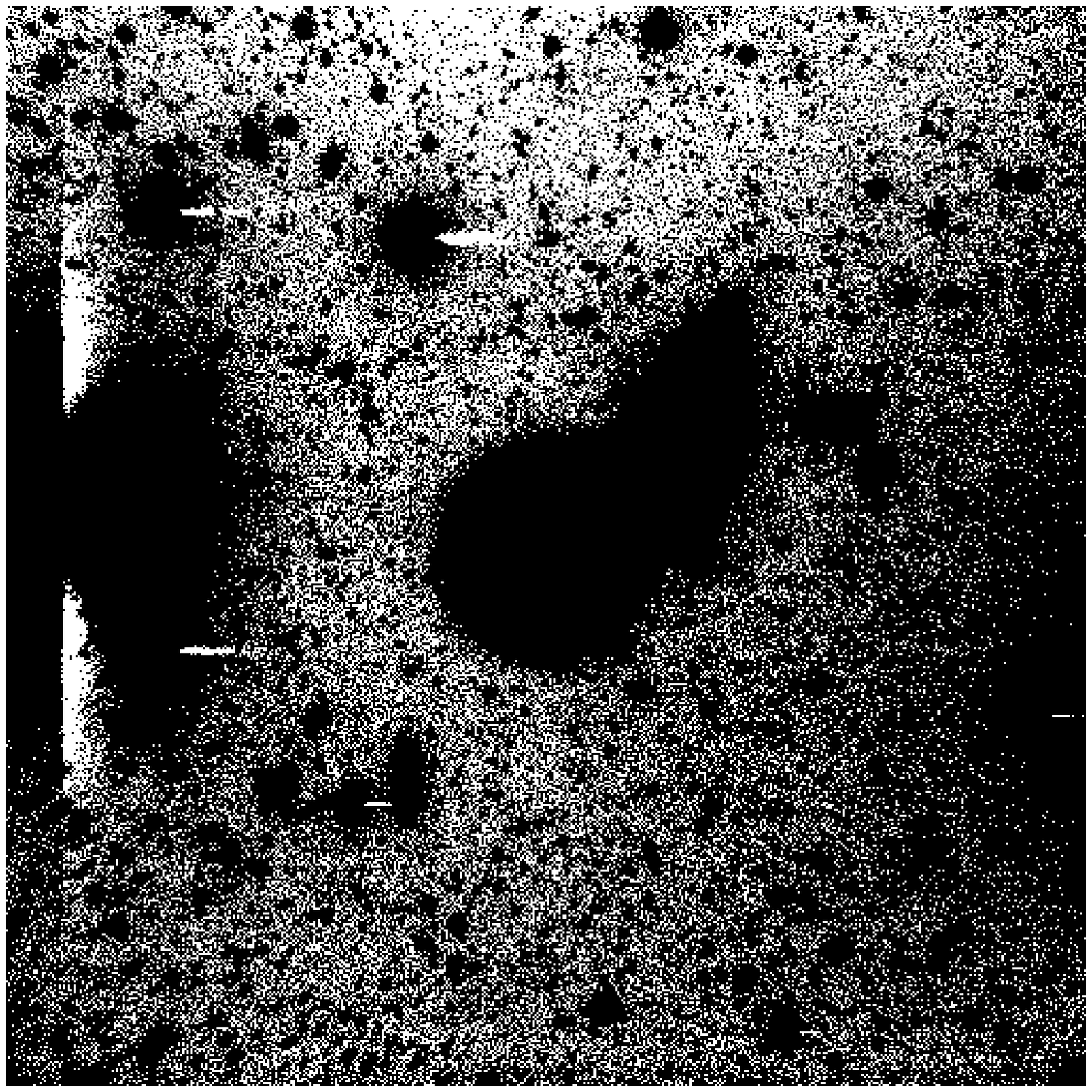}
\caption{Combined images of NGC\,5923, UGC\,9837, and NGC\,5434 displayed with high contrast to highlight the residual background structure. The white tails starting at brighter stars are caused by overexposed pixels. \label{N5923_back} }
\end{figure*}
\subsection{Flatfielding}
\label{flatfielding}
One of the limiting conditions for deep surface photometry 
is the accuracy of the flatfield.
It turned out that both dome-flatfields and twilight-flatfields
are able to remove the pixel-to-pixel variation but on cost of 
a significant large scale gradient remaining on the flatfielded 
science frames.
Depending on the position of the dome slit towards the telescope/camera
system  during the time the twilight-flatfields were taken (90\degr difference  between dusk and dawn), the residual large scale pattern on the flatfielded images changes in the same direction.
Therefore we have to assume that this structure is caused by stray light 
of the illuminated dome during the twilight-flatfielding as well as
the dome-flatfielding procedure.
An additional pupil stop used in the 4th night to minimize this 
stray-light did not solve the problem.
Since we took approximately 20 science images each night with 
significantly dithering across the FOV we have tried to use a nightsky-flat 
median combined from all long-exposure science images during the night
after an additive median scaling and a $3\sigma$ lower and 
$2\sigma$ upper sigma-clipping with {\sl imcombine}.  
It turned out that even with {\sl imcombine's} highly sophisticated 
scaling and rejecting algorithms it is not possible to remove 
all the structure in the combined nightsky-flat. There is
still some residual pattern in the center of the image remaining.
Taking all the 55 science images ($t_{\rm exp}\!>\!600$s) 
from the first three nights within the RR filter together yields 
a decent result for the object residuals and a similar STDDEV compared
to the dome- and sky-flats, but at the expense of some residual underlying 
large scale structure.
Another possibility is to isolate the imposed large scale gradient 
by using dome- or twilight-flats with the {\sl imsurfit} task and 
divide the normalised surface-fit (third order) back into the image again.
This works nicely for the saddle surface introduced by the twilight-flatfields, but not for the ring-like structure of the dome-flatfield.
A major drawback of this method is that one has to ensure that the
object ---compared to the structure of the underlying surface--- is 
excluded from the fit. 
Although it is possible for the face-on galaxies with the {\sl imsurfit} option
'invcirc' to adjust the position of the galaxy center for each fit as well
as a radius, the task fails for the rather large edge-on galaxies chosen.
The overall quality of the combined image relies in this case 
on the quality of the 12 to 17 individual fits to the sky-flatfielded
single images.
In order to minimise the number of crucial fitting processes we finally 
used a third method.
By combining all 55 dome-flatfielded science images with {\sl imcombine}
using a multiplicative exposure time scaling, an additive median scaling, 
and a $3\sigma$ lower and $2\sigma$ upper sigma-clipping, the fairly constant 
(over all nights) dome-flatfield structure is isolated. 
After a median smoothing (15 pixel) to increase the S/N ratio and normalising of the image, this structure is divided back into the dome-flatfield images to yield the final single images without depending on the {\sl imsurfit} quality.
\subsection{Combining}
We test three different ways to combine the individual (up to 
18) object frames. One combining method is a slightly modified version of
the {\sl mosaic} task from the ARNICA\footnote{External IRAF add-on package, 
provided by L.~Hunt, Arcetri (Italy).} 
package, where we introduce the full capabilities of the 
IRAF {\sl imcombine}-task instead of the predefined {\sl combine} 
task using the provided centering routine. 
The {\sl mosaic}-task
allows only full pixel shifts in x- and y-direction without 
rotation or magnification.
In addition, we use the custom written IRAF script 
{\sl gacombine\footnote{Could be provided upon request.}} based on 
the available IRAF centering task {\sl imcntr} together with  IRAFs {\sl geomap/geotran}-tasks to perform the actual matching of all the frames to one reference frame.
We try to use the full transformation with subpixel-shifts, rotation
and distortion of the pixels and a more restricted approach 
using only subpixel-shifts.
The object of comparison reveals that all three methods yields rather similar
results, although the FWHM of selected stars in the {\sl gacombine}d image 
using the full transformation is slightly ($0.1$\arcsec) better.  
To account for possible changes in the overall geometry during the 
rather large ($>3$h) observing period the  {\sl gacombine} task 
with full transformation is chosen to create the final 
combined image.
The {\sl imcombine} task parameters are adjusted to average the 
individual pixels after multiplicative exposure time scaling, 
additive median scaling, and a $3\sigma$-clipping using the 
{\sl ccdclip} algorithm.
The additive median scaling of the original images is essential 
since variation in the sky level (mainly caused by different 
moon positions) during the night are of the order of 0.4 mag.
The final combined images are shown in \fig\ref{N5923_back} with high contrast to highlight the residual background structure.
\subsection{Sky subtraction}
The combined images are visually inspected and foreground stars and galaxies
are masked by hand with the IRAF task {\sl imedit} using  
at least three different radii depending on the object size.
Accurate sky subtraction is crucial for the slope of the derived 
luminosity profiles especially for the low luminosity outer part. 
The actual problem is how and where to determine the single value 
which is called ``the sky value''.
It should be a mean value for all the remaining background pixels on the image still including the galaxy, fainter foreground stars, and remaining background sources. There are several different possibilities to determine such a characteristic value within the available data reduction packages IRAF and MIDAS\footnote{Munich Image Data Analysis System: \newline\hspace*{0.5cm} http://www.eso.org/projects/esomidas/}. They are all aiming to determine either the median (middle value) or the mode (the most probable value) of the pixel distribution.
However, it turns out that neither the 'MIDPT'-value (an estimate of the median of the pixel distribution) nor the 'MODE'-value determined 
with the IRAF {\sl imstat} task\footnote{Note: Lowering the standard 
binwidth of 0.1 improves the quality.} over the whole image  are efficient to determine a reasonable value for the background. Both overestimate the true background significantly.
Therefore we tested three different methods to approximate the most common value ---assumed to be the representative sky value--- within the final image.
\setlength{\leftmargini}{1.0cm}
\begin{enumerate}
\item A mathematical mode (the most frequently occurring) value, determined by eye as the local extremum within the histogram of all pixel values plotted with {\sl imhistogram} 
(\cf\fig\ref{N5923_hist_s}).
\begin{figure}[tb]
\includegraphics[width=6.4cm,angle=270]{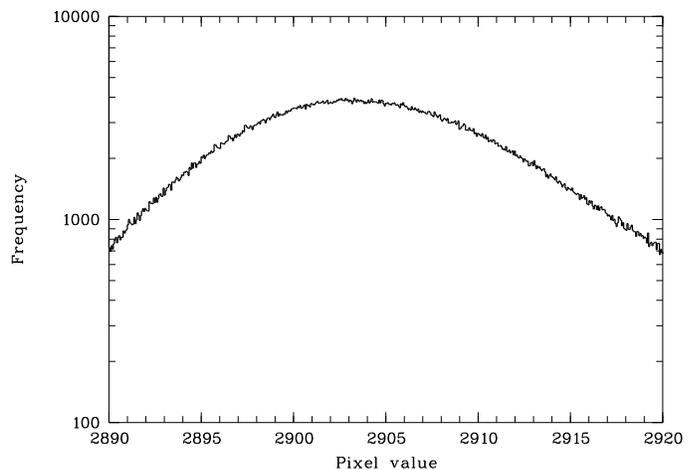}
\caption[]{Histogram of all pixel values in the star-masked image of NGC\,5923 centered on the background peak.
\label{N5923_hist_s} }
\end{figure}
\item The 'MIDPT'-value determined with IRAF's {\sl imexam} on several 
5x5 pixel rectangular regions centered  on  the cursor position
visually pointing to 'empty' sky regions.
\item An exact median determined with the MIDAS task {\sl STATISTICS/IMAGE}, which sorts all pixels of the image by their values.
\end{enumerate}
The problem becomes obvious comparing the different results in the case of 
NGC\,5923: 2903.0, 2901.4, and 2905.2, counts respectively.
In \fig\ref{N5923_sky} the resulting radial profiles are shown using these 
values.
\begin{figure}[tb]
\includegraphics[width=5.5cm,angle=270]{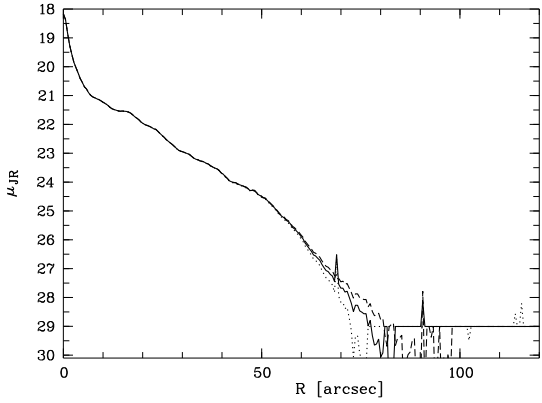}
\caption{Sky subtraction problem: Azimuthally averaged radial profile of NGC\,5923 using the 'empty' region sky-value {\sl (dashed line)}, the histogram method {\sl (solid line)}, and the exact median {\sl (dotted line)}. \label{N5923_sky} }
\end{figure}
The profiles start to differ below a surface brightness of $\mu\!\approx 27.0$\,JR-\magsqarcsec, whereas the actual 1$\sigma$-background fluctuations ---usually used to determine the depth of the image--- reach $\mu\!\approx 29.0$\,JR-\magsqarcsec.
We finally decide to use the histogram method, since this produced the most straight profile beyond $\mu\!\approx 27.0$\,JR-\magsqarcsec without an additional curvature, continuing the undoubtedly linear profile from the break radius to this point. This prevents introducing another slope just at our limiting surface brightness. 
The uncertainty caused by the visual determination is shown in \fig\ref{U09837_skyerror}. In addition to the mean sky the two associated justifiable extremes are plotted.
This discrepancy slightly affects the determined slope of the outer profile.
\begin{figure}[tb]
\includegraphics[width=5.5cm,angle=270]{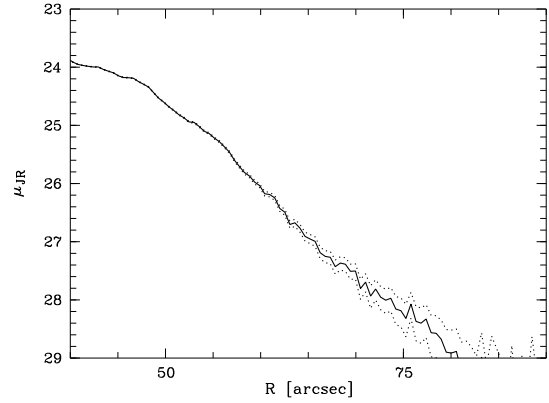}
\caption{Sky subtraction problem: Azimuthally averaged radial profile using different values obtained for UGC\,9837 with the histogram method {\sl (\cf text)}. \label{U09837_skyerror} }
\end{figure}
In this case the profiles start to deviate significantly below $\mu\!\approx 27.2$\,JR-\magsqarcsec which marks the final accuracy of our photometry.
\subsection{Photometric calibration}
Several Landolt standard fields \cite[]{landolt} have been observed each night following at least one over several airmasses to calculate the extinction coefficients.
The calibration coefficients are determined according to 
following equations:
\beas
jb-JB &=&c_{0,{\rm JB}} +c_{1,{\rm JB}}\cdot X  +c_{2,{\rm JB}}\cdot (JB-JR) \\
jr-JR &=&c_{0,{\rm JR}} +c_{1,{\rm JR}}\cdot X  +c_{2,{\rm JR}}\cdot (JB-JR) 
\eeas
\begin{table*}
\begin{center}
\begin{tabular}{l c c c c}
\hline
\rule[+0.4cm]{0mm}{0.0cm}
Date & Band & Zero point & Airmass & Colour \\
 &  & $c_0$ & coef. $c_1$ & coef. $c_2$ \\[+0.05cm]
\hline\hline 
N1: 26.05.01&JR&$-23.993 \pm 0.098$&$0.178 \pm 0.035$&$+0.030 \pm 0.095$\\
N2: 27.05.01&JR&$-23.973 \pm 0.134$&$0.103 \pm 0.085$&$+0.009 \pm 0.095$\\
N3: 28.05.01&JR&$-23.987 \pm 0.012$&$0.096 \pm 0.060$&$+0.004 \pm 0.012$\\
N4: 29.05.01&JR&$-23.675 \pm 0.302$& --- & ---  \\
N5: 30.05.01&JR&$-23.986 \pm 0.015$&$0.103 \pm 0.008$&$+0.003 \pm 0.015$\\
\hline \\
\end{tabular}
\caption{Photometric calibration coefficients used for each night. \label{calCA0105}}
\end{center}
\end{table*}
For the fourth night no reasonable solution is found and only a zero point could be determined. The applied coefficients are shown in Tab.~\ref{calCA0105}
The fields were observed in Johnson R and B as well as the R\"oser R
filter to quantify the difference between RR and JR.
We do not find a systematic correlation of the RR-magnitudes and
JR-magnitudes with star colour or airmass. Therefore, we assume
in the following the mean difference of $m_{\rm JR}\!-\!m_{\rm RR}\eq0.35\pm0.02$\,mag.
\subsection{Polar transformation}
\label{polar}
\begin{figure}[tb]
\includegraphics[width=4.7cm,angle=270]{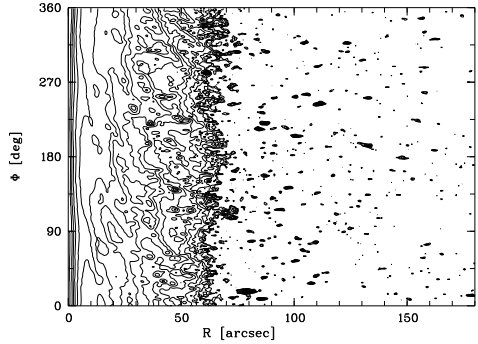}
\caption{Contour map  ($\mu_{\rm JR}$)  of NGC\,5923 after transformation into polar coordinates between 26.4 and 17.9 in 0.5\,mag steps. \label{N5923_polar}} 
\end{figure}

\begin{figure}
\includegraphics[width=5.5cm,angle=270]{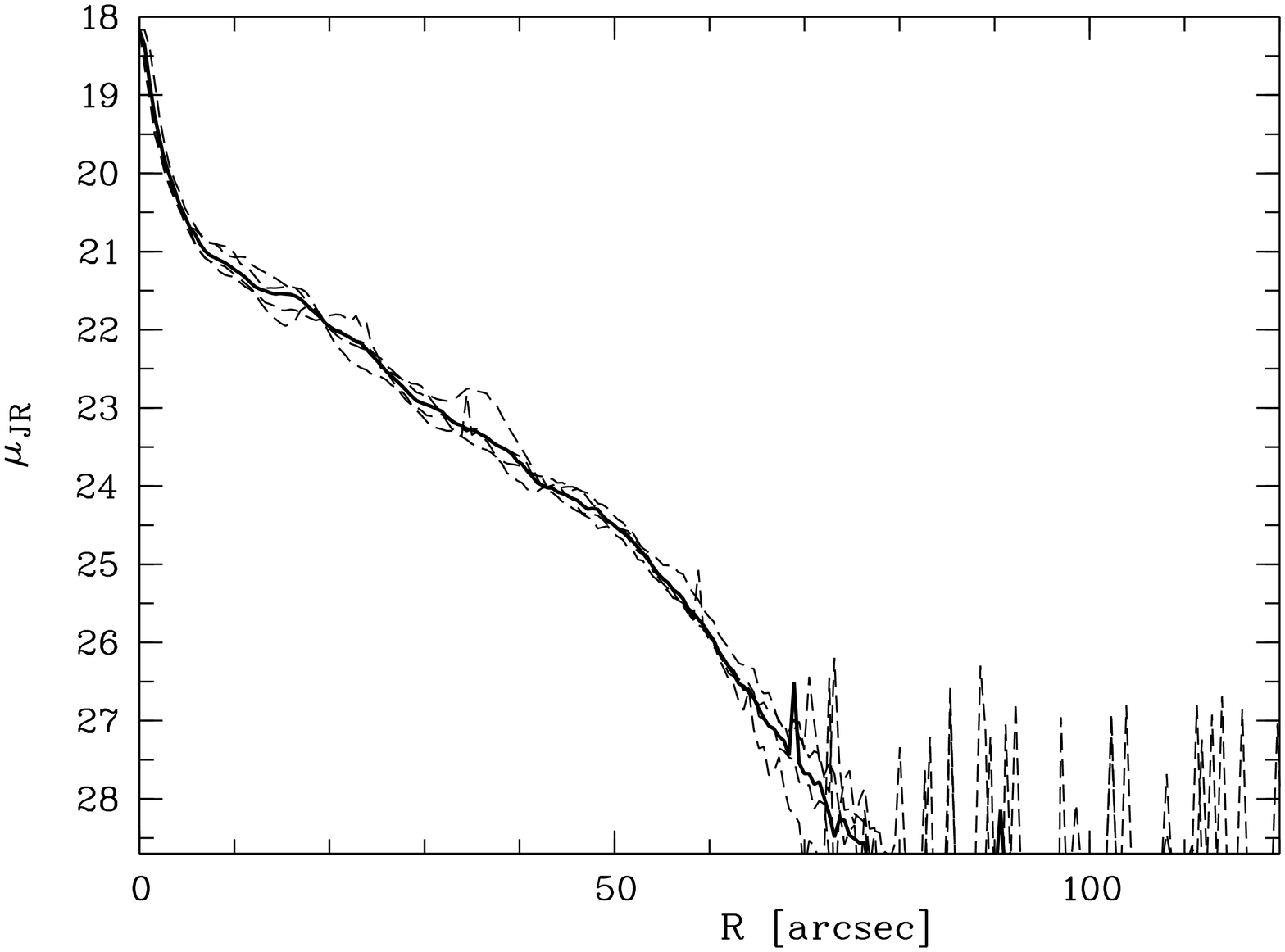} \\
\includegraphics[width=5.5cm,angle=270]{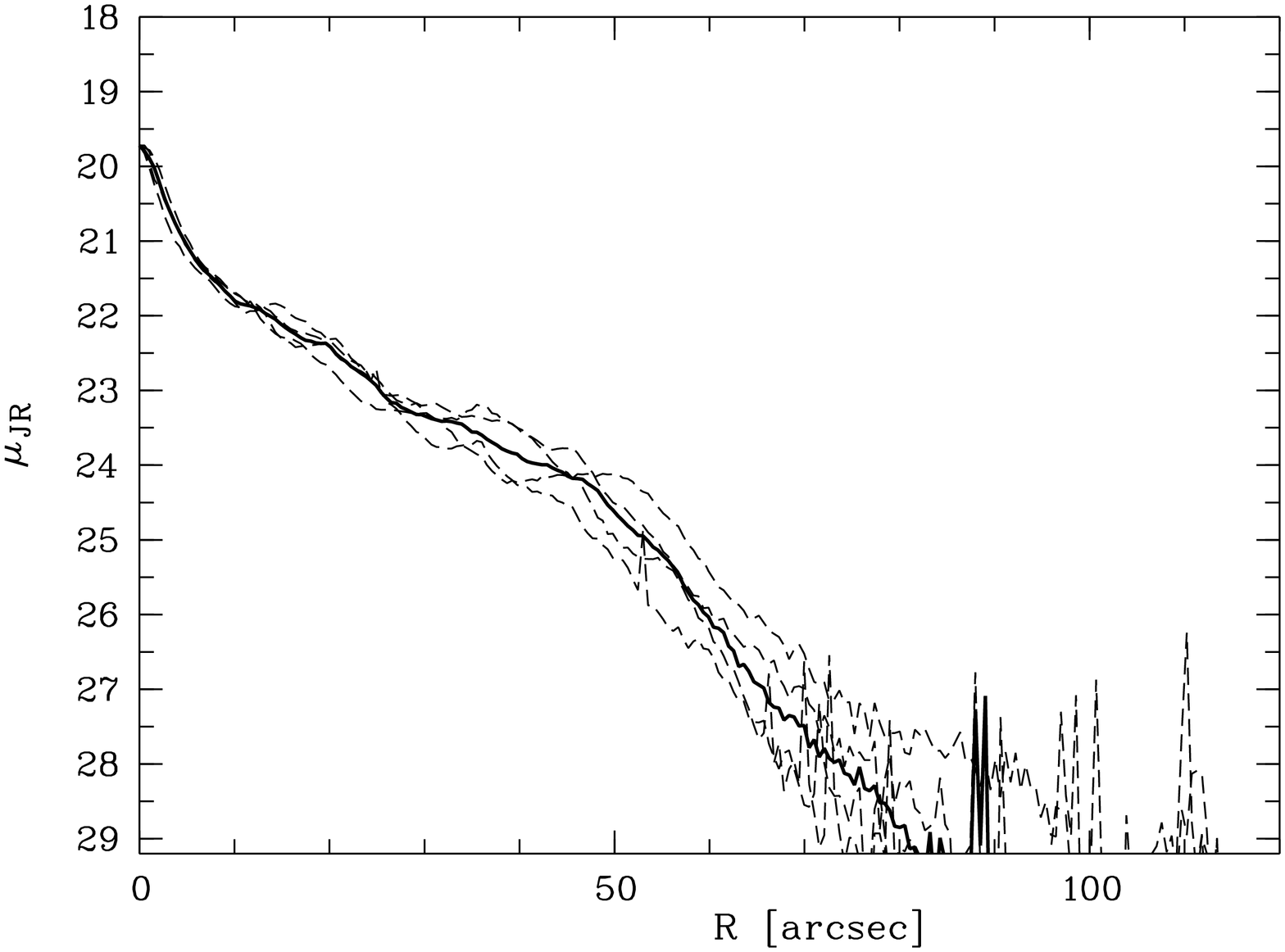} \\
\includegraphics[width=5.5cm,angle=270]{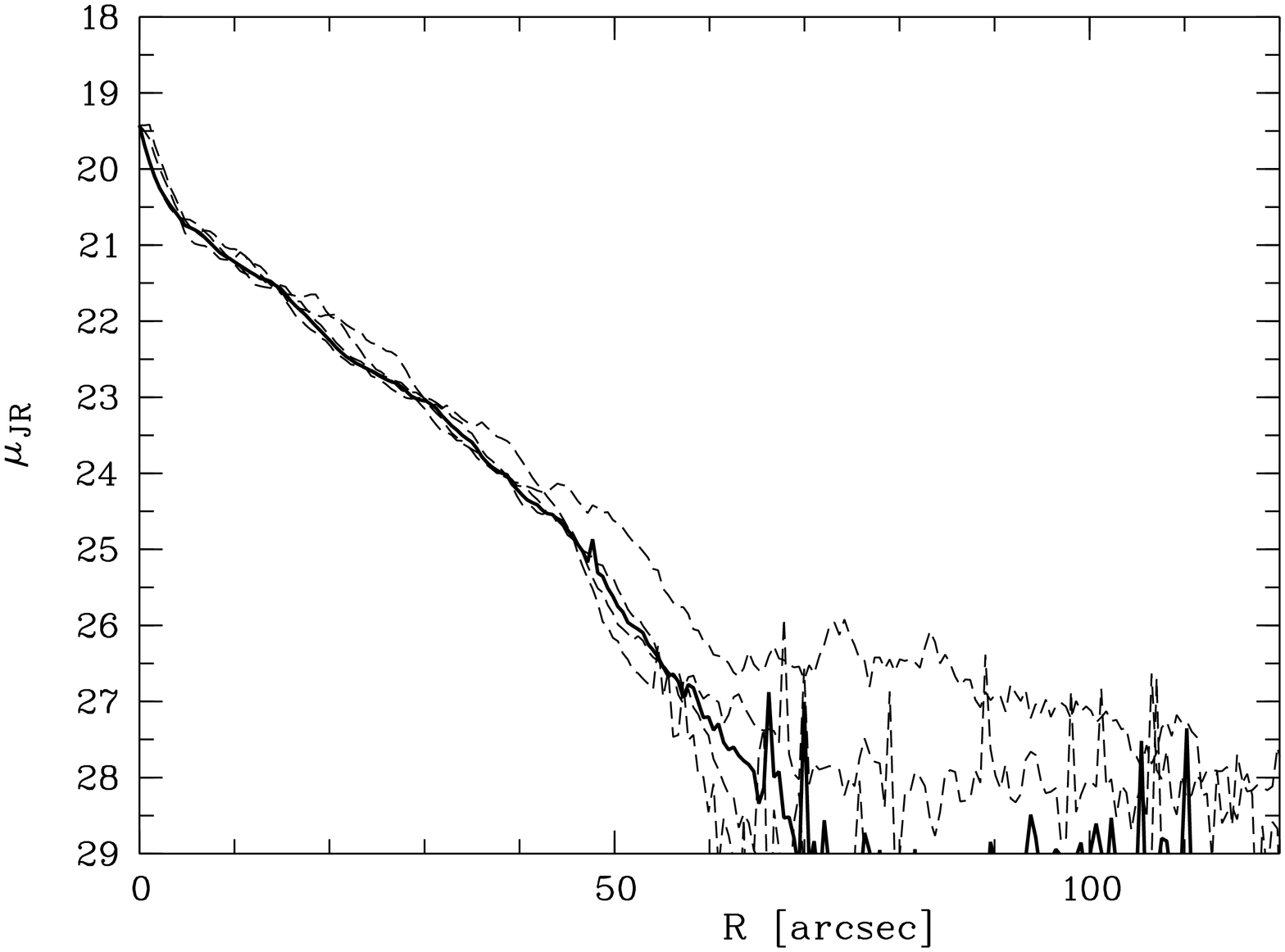}
\caption[test]{Combined 360\degr-segment {\sl (thick solid line)} together with the four 90\degr-segments {\sl (thin dashed lines)} Radial profiles of NGC\,5923 {\sl (upper panel)}, UGC\,9837 {\sl (middle panel)}, and NGC\,5434 (only 180\degr) {\sl (lower panel)}. \label{comsegm} }
\end{figure}

Using the first order symmetry of the chosen ideal face-on galaxies 
an azimuthal averaging within predefined segments increases the
signal-to-noise ratio and allows to follow a smooth profile down to 
the limits of the photometry.
In order to independently pick up the size of these segments 
the whole image is transformed into polar coordinates using 
an already existing C-program\footnote{Courtesy L.~Schmidtobreick.}.
The polar image of NGC\,5923 is shown in \fig\ref{N5923_polar}.
By combining the individual rows ---split with the IRAF task {\sl imslice}
and apply the powerful rejecting and combining algorithms of {\sl imcombine}---
we are able to mimic an azimuthally smoothing within segments of the 
galaxy disk and to reduce the influence from individual \htwo regions.
NIR images would be more suited to study the underlying mass distrubution, although currently probably infeasible to these depths. The optical passband used is therefore contaminated by star-formation in and outside the spiral arms.
These intrinsic deviations within the galaxy's structure from azimuthal symmetry are obviously present (\cf individual profiles in \fig\ref{comsegm}).
However, our azimuthally smoothing by combining the individual sliced rows does not significantly change the inner or blur the outer slope it rather averages the individual slopes in a reasonable manner, in contrast to the statement of \cite{vdk88}.
\section{Analysis \& Results}
\subsection{Derived parameters}
\begin{figure}
\includegraphics[width=5.5cm,angle=270]{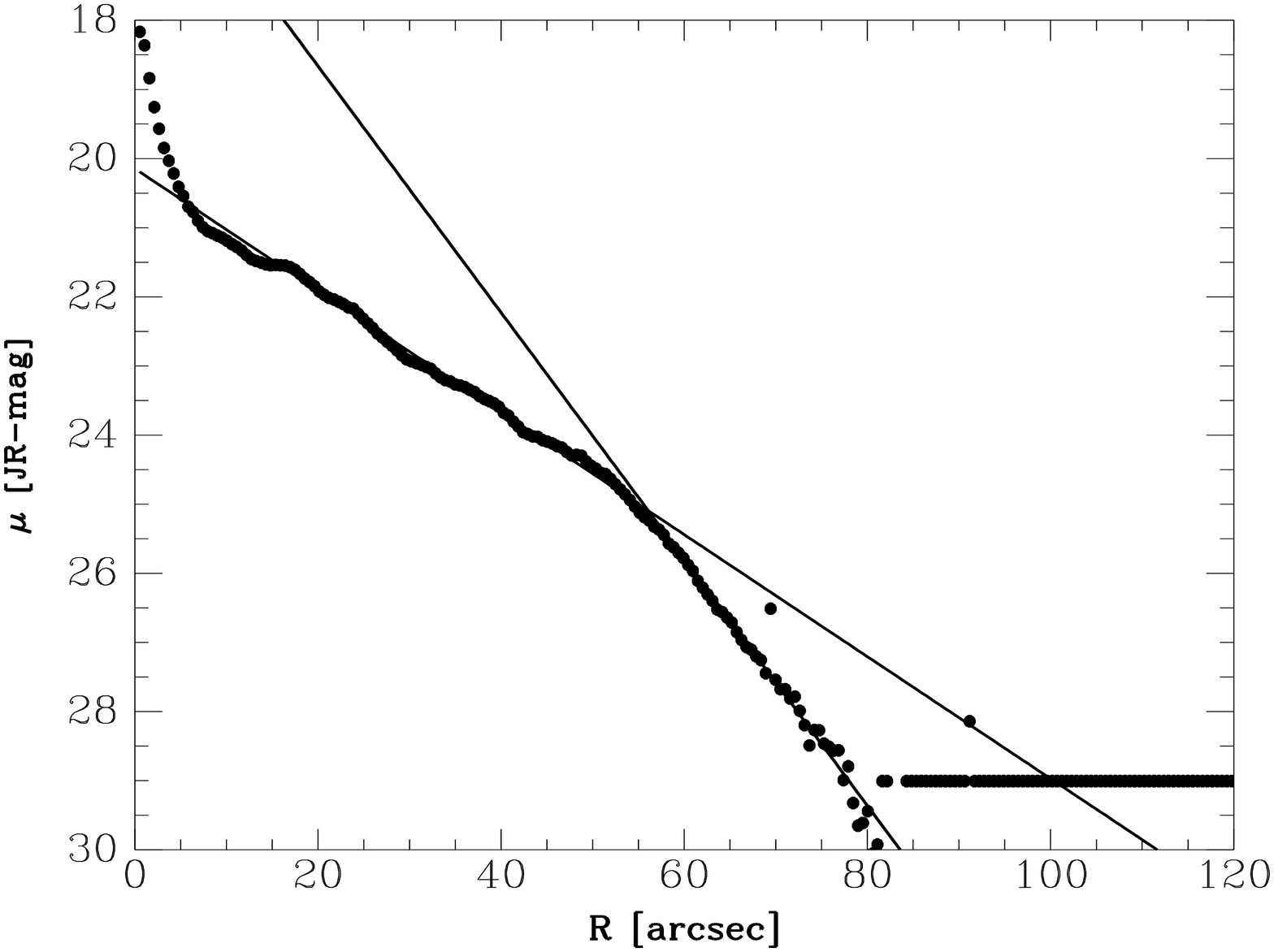} \\
\includegraphics[width=5.5cm,angle=270]{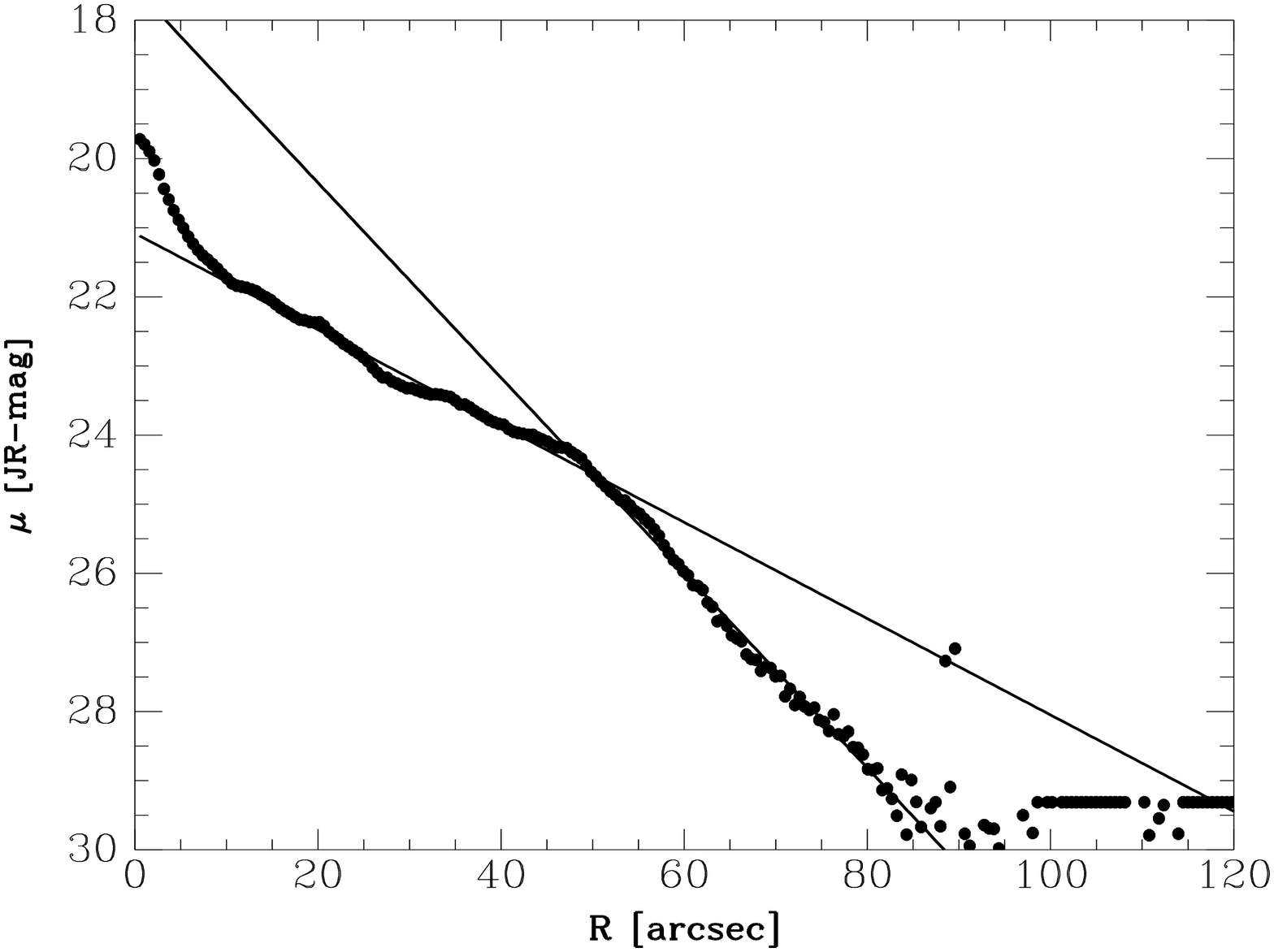} \\
\includegraphics[width=5.5cm,angle=270]{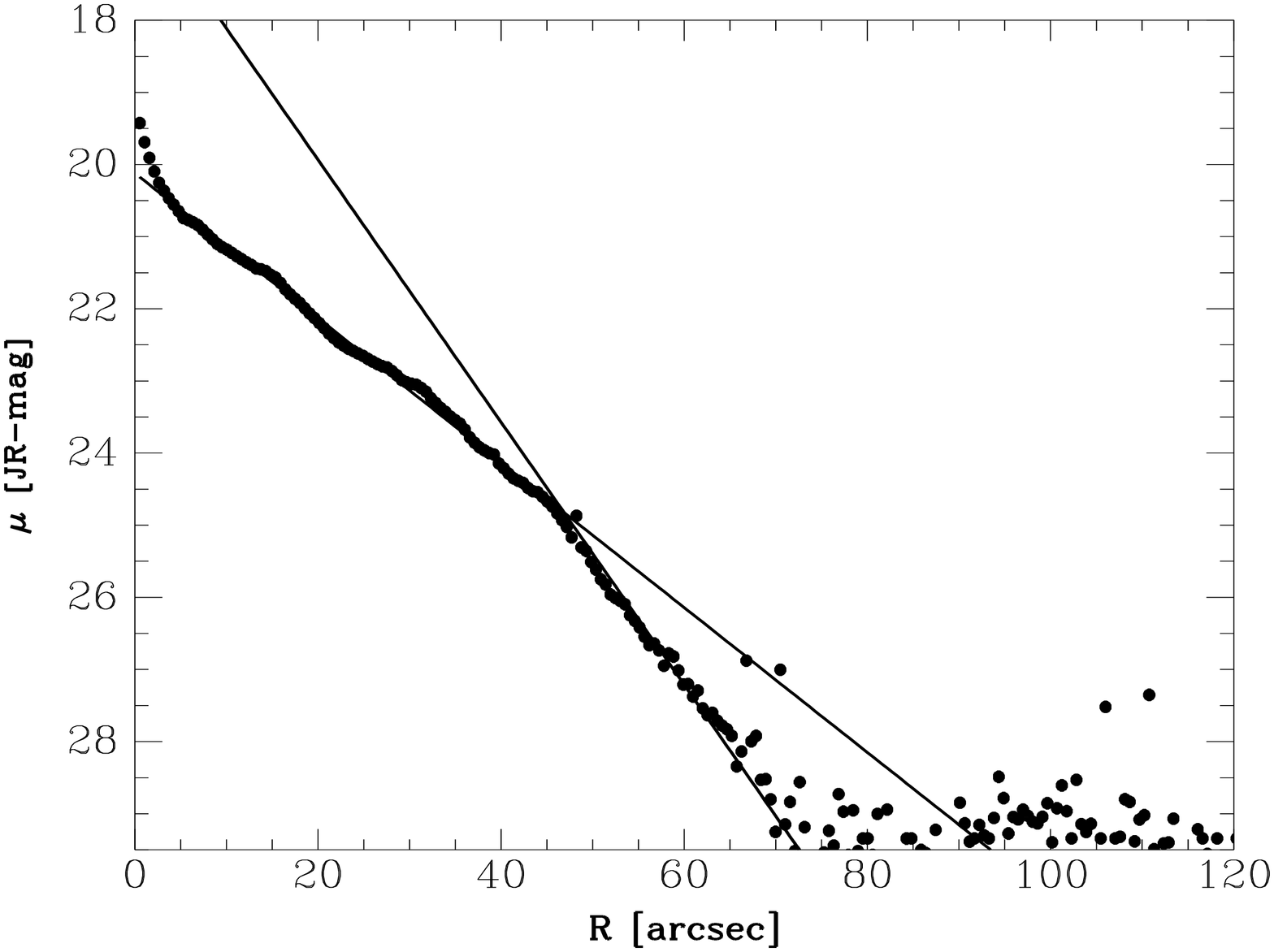} \\
\caption[]{One dimensional exponential fits to the two distinct parts separately overplotted on the radial profiles: NGC\,5923 {\sl (upper panel)}, UGC\,9837 {\sl (middle panel)}, and NGC\,5434 {\sl (lower panel)}. \label{1dexpfit}}
\end{figure}

We limit the following analysis of the radial profiles to the 
360\degr\,averaged profiles, except for NGC\,5434 where only the 180\degr\,
profile, avoiding the edge-on companion, are taken into account.
The final combined profiles together with all four 90\degr\,segments obtained from the polar transformed image (see Sect.~\ref{polar}) are shown in \fig\ref{comsegm}.
In all three cases we find a common behaviour.
The radial surface brightness profile is best described by a two-slope model, characterised by an inner and outer exponential scalelength separated at a rather sharp so called {\it break radius} $R_{\rm br}$. 
The inner approximately $8\!-\!10$ arcsec are dominated by a bulge component followed by a rather well described first exponential decline of 3.3 mag within $\approx\!41$\arcsec and a second also quite well described second exponential decline of 4.4\,mag (down to the observing limit) within $\approx\!27$\arcsec.
Taking into account the estimated distance this translates into an inner approximately $2.6$\,kpc wide region dominated by the bulge component followed by a first exponential decline of radially rather different sizes for the three galaxies (NGC\,5434: $\Delta R\!\approx\!7.7$\,kpc, UGC\,9837: $\Delta R\!\approx\!12.1$\,kpc, NGC\,5923: $\Delta R\!\approx\!18.1$\,kpc) and a second slope within $\Delta R\!\approx\!7.8$\,kpc limited by the outermost measured point. 
The inner and outer slope are determined with an 1D exponential fit of the form $I(R)\!=\!I_0\exp(-R/h)$ to the combined radial surface brightness profile using the IRAF task {\sl tlinear}. The inner boundary is defined by the bulge light and the fit is performed down to $\mu\!\approx\!28.5$\,JR-\magsqarcsec while the break radius is determined by eye.
The resulting fits are shown in \fig\ref{1dexpfit}.
The resulting values are shown in \tab\ref{F2results}, for the break radius $R_{\rm br}$\,{\scriptsize{\it (4)}} in units of arcsec, kpc (with the estimated distance $D$ {\scriptsize{\it (2)}}), and in units of the inner scalelength $h_{\rm in}$\,{\scriptsize{\it (5)}}, the surface brightness $\mu_{\rm br}$\,{\scriptsize{\it (3)}}) at the position of the break radius, as well as the ratio of inner to outer scalelength $h_{\rm in}/h_{\rm out}$ {\scriptsize{\it (7)}}).
The total sizes defined as the last measured point of their disks are:
30.5\,kpc (NGC\,5923), 16.0\,kpc (UGC\,9837), and 22.3\,kpc (NGC\,5434).
Although being probably an interacting system the derived values for NGC\,5434 are in no way exceptional compared to the other two galaxies.
\subsection{Comparison with the literature}
\begin{table*}
\begin{center}
\begin{tabular}{l c c c r@{.}l c r c c r c c r c}
\hline
\rule[+0.4cm]{0mm}{0.0cm}
Galaxy
&D
&$\mu_{\rm br}$ 
&\multicolumn{4}{c}{$R_{\rm br}$}
&
&\multicolumn{2}{c}{$h_{\rm in}$}
&
&\multicolumn{2}{c}{$h_{\rm out}$}
&
&$h_{\rm in}/h_{\rm out}$ \\
\cline{4-7}\cline{9-10}\cline{12-13} \\[-0.15cm]
&\raisebox{0.0ex}{[Mpc]}
&\raisebox{0.0ex}{[JR-\magsqarcsec]}
&\raisebox{0.0ex}{[\,\arcsec]}
&\multicolumn{2}{c}{\raisebox{0.0ex}{[kpc]}}
&\raisebox{0.0ex}{$[h_{\rm in}]$}
&
&\raisebox{0.0ex}{[\,\arcsec]}
&\raisebox{0.0ex}{[kpc]}
&
&\raisebox{0.0ex}{[\,\arcsec]}
&\raisebox{0.0ex}{[kpc]}
& \\
\rule[-3mm]{0mm}{5mm}{\scriptsize{\raisebox{-0.7ex}{\it (1)}}}
&{\scriptsize{\raisebox{-0.7ex}{\it (2)}}}
&{\scriptsize{\raisebox{-0.7ex}{\it (3)}}}
&\multicolumn{4}{c}{\scriptsize{\raisebox{-0.7ex}{\it (4)}}}
&
&\multicolumn{2}{c}{\scriptsize{\raisebox{-0.7ex}{\it (5)}}}
&
&\multicolumn{2}{c}{\scriptsize{\raisebox{-0.7ex}{\it (6)}}}
&
&{\scriptsize{\raisebox{-0.7ex}{\it (7)}}} \\
\hline \hline 
&&&& \multicolumn{2}{}&&&&&&&&\\[-0.3cm]
NGC\,5923 &80.7& 24.7  &$54.0_{-2}^{+3}$&21&$1_{20.3}^{22.3}$&$4.30_{4.13}^{4.63}$& &$12.6_{12.6}^{12.3}$&$4.9_{4.9}^{4.8}$& &$6.2_{6.3}^{6.1}$&$2.4_{2.5}^{2.4}$& &$2.0_{2.0}^{2.0}$\\[0.1cm]
UGC\,9837 &40.8& 24.4  &$49.0_{-5}^{+5}$& 9&$7_{\,\;8.7}^{10.7}$&$3.05_{2.86}^{3.47}$ & &$16.1_{15.4}^{15.6}$&$3.2_{3.0}^{3.1}$ & &$7.6_{7.8}^{7.7}$&$1.5_{1.5}^{1.5}$& &$2.1_{2.0}^{2.0}$\\[0.1cm]
NGC\,5434 &65.8& 24.7  &$46.0_{-3}^{+1}$&14&$7_{13.7}^{15.0}$&$4.23_{3.92}^{4.35} $& &$10.9_{11.0}^{10.8}$&$3.5_{3.5}^{3.5} $& &$6.0_{6.1}^{6.0}$&$1.9_{2.0}^{1.9}$ & &$1.8_{1.8}^{1.8}$\\[0.1cm]
\hline 
mean     &    & 24.6 &    & \multicolumn{2}{c}{} &$3.9\pm0.7$  & &    &    & &   &    & &$2.0\pm0.2$\\ 
\hline 
\hline
\end{tabular}
\caption{Results from the 1D exponential fitting of the radial surface brightness profiles using fixed inner and outer boundaries and tenably varying the break radius around the mean value.  \label{F2results}}
\end{center}
\end{table*}
The observed smooth nature of the radial truncation is consistent with earlier work on edge-on galaxies. Already early photographic profiles fitted by \cite{vdk81a} (\eg their Fig.~9 for NGC\,5907) show the smooth nature of the truncation and reveal that their sharply-truncated model explicitly overpredicts the luminosity profile at the truncation region.
A similar behaviour is found in the recent work of \cite{pohlen2} where they fit the sharply-truncated model of \cite{vdk81a} to CCD observations.
The derived two-slope model consisting of two exponentials separated at a break radius is also discussed by \cite{naeslund} and \cite{byun}.
By analysing a deep optical image of NGC\,4565 \cite{naeslund} find a similar two-slope shape for their major axis profile (\cf their Fig.~2). They state that exponentials also provide a good solution at both sides beyond the break radius and that there is no clear hint in their data of any further change in the disk scalelength at surface brightnesses lower than $\mu_V\eq28$\,\magsqarcsec.
The same experience is made by \cite{byun} in the case of IC\,5249. He also discovers that the major axis of IC\,5249 demonstrates a peculiar two-component structure with an exponential inner disk, where the change of surface brightness is slow and also an exponential outer disk with a steep change of surface brightness. He concludes that this structure cannot be caused by internal absorption but must represent the intrinsic stellar distribution.
Prior to this already \cite{jensen} for NGC\,4565 and \cite{sasaki} for NGC\,5907 fitted additional exponential profiles to the truncation regions.
A different approach is outlined by \cite{floridoNIR}. They use a model independent deprojection method to measure and analyse the truncation curve which is the difference between the measured profile and an infinitely exponential decline. 
In the case of face-on galaxies \cite{vdkshos82} measure for NGC\,3938 a quite similar radial profile (\cf their \fig10) with a break radius at 
$R_{\rm br}\eq4.2\!\cdot\!h_{\rm in}$
($\mu_{\rm br}\!\approx\!24.5$\,F-\magsqarcsec)
and a value for the ratio of inner and outer scalelength of 
$h_{\rm in}/h_{\rm out}\eq1.6$ which is consistent for the 
occurrence of the break radius and only somewhat smaller for the 
scalelength ratio compared to our results.
Using the radial velocity of 1011\,\kms from 
LEDA\footnote{Lyon/Meudon Extragalactic Database (LEDA).}
corrected for the Local Group infall onto Virgo and a Hubble constant of $
H_{0}\!=\!72$\,\kms\,Mpc$^{-1}$ \cite[]{hst_h0}, we estimate
the distance of NGC\,3938 to 14.0\,Mpc. 
Thereby we recompute the linear parameters and recover a rather low 
inner scalelength of 2.5\,kpc, a more typical outer scalelength of 
1.5\,kpc, and a break radius at 10.2\,kpc which are all similar 
to our values.
In contrast to \cite{vdkshos82} we do not regard our derived 
outer scalelengths as upper limits of the true scalelength.
They assume this upper limits, due to the azimuthally averaged 
deviations from the pure circularity of the isophotes.
It is obvious by comparing the $360\degr$ averaged profile to the individual $90\degr$ wide strips (\cf Fig.~\ref{comsegm}) that the azimuthally averaging does not significantly change the measured outer scalelength. There are of course variations between the individual profiles due to spiral arms, prominent \htwo regions, and intrinsic asymmetries, but judging from the $360\degr$ averaged profile it seems to be a reasonable mean representation.
The radial surface brightness profile of NGC\,628 presented by 
\cite{shos} also fits quite well with our results (\cf their \fig13).
They find a break radius at precisely $R_{\rm br}\eq3.1 h_{\rm in}$ ($\mu_{\rm br}\!\approx\!23.9$\,F-\magsqarcsec), although quoting to be at ''$\sim 4$ scalelength'',  and a value for the ratio of inner and outer scalelength of $h_{\rm in}/h_{\rm out}\eq2.6$, together with an inner scalelength of $4.0$\,kpc and an outer of $\approx\!1.5$\,kpc.  
Comparing our results with other published radial profiles of NGC\,628
\citep{natali} also supports the two-slope model.
According to our mean value of $\rbrdhin\eq3.9\pm0.7$ and using their derived scalelength $h_{\rm R-band}\eq3.3$ we predict a break in the profile to be visible at $13\pm2$\,kpc.
Assuming their profiles to be reliable out to 16\,kpc (\cf their Fig.~8 and Fig.~9) we find a clear break in the B, V, R and I profiles with similar shape compared to our three galaxies at 14\,kpc equivalent to $\rbrdhin \eq 4.2$.
Therefore their results fit well to our results for NGC\,5923 and NGC\,5434.
Unfortunately their colour profiles just stop at this radial distance and we are not able to find a similar break in the colour profiles which would possibly constrain the physical hypothesis for the origin of these truncations.
\subsection{Constraints for the origin of truncations}
In order to search for a possible correlation of the 
break radius and the presence of spiral arm structure 
we visually determined the extent of the spiral arms 
for the target galaxies.
It is well known that the spiral structure is constrained between 
the inner Lindblad resonance (ILR) and the outer Lindblad resonance (OLR) 
\cite[eg.][]{canzian}.
Unfortunately all three galaxies are not well defined 
grand-design spirals. 
NGC\,5923 and NGC\,5434 show nearly similar behaviour starting
with a two-arm spiral which is splitting up towards the outer
parts in at least four traceable arms.
UGC\,9837 is even more complex. In the inner part starting 
with a two-arm spiral, it looks rather flocculent in the 
outer parts. 
The estimated extent of the spiral arms are $60.4$\arcsec (NGC\,5923), 
$58.8$\arcsec (UGC\,9837), and $46.1$\arcsec (NGC\,5434).
In the case of NGC\,5434 this fits well with the derived break 
radius, although for NGC\,5923 it is $\approx\!6$\arcsec (2.5\,kpc) and for 
UGC\,9837 $\approx\!10$\arcsec (1.9\,kpc) larger than \rbr.
However, \cite{canzian} already points out that it is unknown how often 
spiral structure remains relatively bright, or even optically detectable,
at radii approaching the OLR. 
He derives for his sample of face-on to moderately inclined galaxies 
a large scatter of the spiral arm surface brightness at their outer 
extent, ranging from $\mu \eq 22.6 - 26.5$\,V-\magsqarcsec, compared  
to our mean value of $\mubr\eq 24.6$\,R-\magsqarcsec.
Kennicutts study on the dependence of the star formation rate (SFR) on the density and dynamics of the interstellar gas \cite[]{kenni} proposes a dynamically critical gas density acting as a star formation threshold. Beyond this critical radius the star formation should be inhibited and if this persists for sufficient time, it should introduce a visible turnover or truncation of the observed stellar luminosity profile at that radius. These critical radii for star formation are measured by prominent breaks in the \halpha surface-brightness profiles \citep[cf.][]{martin+}.
\cite{lelievre} discover that the \halpha surface brightness profiles in the case of NGC\,628 diminish monotonically as a function of galactocentric distance, dropping dramatically at $R\!\sim\!20$\,kpc. 
This fall-off in $\Sigma_{\rm H\,\alpha}$ near the edge of the optical disk
(defined by the 25th B-band isophote) is also observed by \cite{ferg} 
for NGC\,628 and also NGC\,6946. 
\cite{lelievre} conclude that the standard Schmidt law 'fails' at this 
radius equivalent to a gas surface density of 
$\approx\!4$ M$_{\odot}\,$pc$^{-2}$, although they find additional 
\htwo regions out to 27\,kpc, and an \hone disk extending even much further.
However, \cite{lelievre} and \cite{ferg} do not compare their results with {\bf two profiles} tracing the underlying old stellar population which are present in the literature in the case of NGC\,628.
\cite{shos} locate a break radius for NGC\,628 at 12.4\,kpc ($\equiv\!12.9$\,kpc for $D\eq10.4$\,Mpc), whereas we deduce 14\,kpc from \cite{natali}.
Compared to the '\halpha -truncation' at $R\!\approx\!20$\,kpc of \cite{lelievre} optical breaks at 13\,kpc or 14\,kpc clearly rule out a one-to-one correlation.
Nevertheless, at 14\,kpc a small plateau-like feature  starts in the 
averaged \halpha surface brightness \cite[Fig.~3 of][]{lelievre}.
In addition, using the data for NGC\,628 of \cite{ferg} we deduce a clear break in the \halpha profile at $0.8 R_{25}$ (\cf their Fig.~2). With their distance of 10.7\,Mpc this is equal to $\approx\!13$\,kpc ($\equiv\!12.6$\,kpc for $D\eq10.4$\,Mpc) and therefore similar to the break found by \cite{shos} and the one deduced from \cite{natali}.
\section{Conclusions}
Our deep optical imaging of three face-on galaxies confirms the finding of \cite{naeslund}, \cite{byun}, \cite{dgkrwe}, and \cite{pohlenphd} that the observed truncations of galactic stellar disks are not infinitesimally sharp but better described by an additional exponential decline.
Since face-on scalelengths are not influenced by line-of-sight integration problems as in the case for edge-on galaxies (so far investigated), we are able to quantitatively introduce a new two-slope model.
We find that the radial surface brightness profiles are best represented by a two-slope exponential profile, characterised by an inner \hin and outer \hout scalelength separated at a break radius \rbr, rather than the previously assumed sharply-truncated exponential model proposed by \cite{vdk81a} and used for example in the study of \cite{pohlen2}.
The mean value for the distance independent ratio of break radius to inner scalelength which marks the start of the truncation region is $\rbrdhin\!=\!3.9\pm0.7$. 
Although larger statistics are also definitely needed, the lack of any conspicuity in the parameters for NGC\,5434 with its close physical companion points against an interaction hypothesis for explaining stellar truncations.
In addition, we could not correlate undoubtedly the end of the visual spiral arms with the occurrence of the break radii, which may be a hint relating the truncation phenomenon to star formation properties. Similar deep imaging of grand-design spirals are needed to clarify this hypothesis.  
Since we do not have deep \halpha maps for our galaxies we have searched the literature for signs of break radii in published surface brightness profiles and have compared them in the case of NGC\,628 with already available \halpha\,data.
Although not conclusive in this case an \halpha\,follow up of the three observed face-on galaxies will be a promising project to determine the origin for the observed break radii (or smooth truncations) and address the star formation threshold hypothesis by \cite{kenni}.
However, the detailed quantitative structure of the {\it smoothness} of the observed truncations presented here will set strong constraints for present and future theoretical models trying to explain this observational phenomenon.
%
%
\begin{acknowledgements}
      Part of this work was supported by the German
      \emph{Deut\-sche For\-schungs\-ge\-mein\-schaft, DFG}.
We would like to thank the referee, Dr. E. Battaner, for his useful suggestions to improve this paper.
This research has made use of the NASA/IPAC Extragalactic Database (NED) which is operated by the Jet Propulsion Laboratory, California Institute of Technology, under contract with the National Aeronautics and Space Administration. It also uses the Digitized Sky Survey (DSS) based on photographic data obtained using Oschin Schmidt Telescope on Palomar Mountain and The UK Schmidt Telescope and produced at the Space Telescope Science Institute. 
\end{acknowledgements}
\end{document}